\documentclass[12pt,a4paper]{article}
\usepackage{jheppub}

\usepackage{color}
\usepackage{amsmath}

\usepackage{verbatim}   
\usepackage{subfigure}  
\usepackage{amsfonts}
\usepackage{amssymb}
\usepackage{mathrsfs}
\usepackage{graphicx}
\usepackage{slashed}
\usepackage{amsthm}

\usepackage{graphicx}
\usepackage{dcolumn}
\usepackage{bm}

\usepackage{color}
\usepackage{amsmath}

\usepackage{verbatim}   
\usepackage{subfigure}  
\usepackage{amsfonts}
\usepackage{amssymb}
\usepackage{mathrsfs}
\usepackage{graphicx}
\usepackage{slashed}
\usepackage{amsthm}

\def\inv{^{\raise.15ex\hbox{${\scriptscriptstyle -}$}\kern-.05em 1}}
\def\lbar{{\lower.35ex\hbox{$\mathchar'26$}\mkern-10mu\lambda}} 

\let\p=\partial
\let\<=\langle
\let\>=\rangle

\let\+=\uparrow

\let\r=\right

\def\OO{\mathcal{O}}

\theoremstyle{definition}

\begin{document}

\title{Heterotic Moduli Stabilisation and Non-Supersymmetric Vacua}

\author[a]{Andre Lukas,}
\author[b]{Zygmunt Lalak}
\author[c,d,e]{and Eirik E.~Svanes}

\affiliation[a]{Rudolf Peierls Centre for Theoretical Physics, University of Oxford\\ 
1 Keble Road, Oxford OX1 3NP, UK}
\affiliation[b]{Institute of Theoretical Physics, Faculty of Physics,University of Warsaw ul. Pasteura 5, 02-093 Warsaw, Poland}
\affiliation[c]{Sorbonne Universit\'es, UPMC Univ Paris 06, UMR 7589, LPTHE, F-75005, Paris, France}
\affiliation[d]{CNRS, UMR 7589, LPTHE, F-75005, Paris, France}
\affiliation[e]{Sorbonne Universit\'es, Institut Lagrange de Paris, 98 bis Bd Arago, 75014 Paris, France\\}

\emailAdd{lukas@physics.ox.ac.uk}
\emailAdd{zygmunt.lalak@fuw.edu.pl}
\emailAdd{esvanes@lpthe.jussieu.fr}

\null\vskip10pt

\date{\today}

\abstract{We study moduli stabilisation in four-dimensional $N=1$ supergravity theories which originate from compactifications of the heterotic string on certain manifolds with $SU(3)$ structure. These theories have a non-trivial superpotential generated from geometric flux and, in general, D-terms associated to anomalous $U(1)$ symmetries. We show that, at the perturbative level, there are no supersymmetry preserving vacua. However, subject to a certain technical condition on the D-terms which aligns the extrema of the F-term and D-term potentials, $\p_iV_F=\p_iV_D=0$, we find at the perturbative level analytic stable AdS vacua which break supersymmetry. As a result, all T-moduli and the dilaton are stabilised perturbatively with supersymmetry broken at a high scale. We also show numerically that similar vacua can be found when the technical condition on the D-term is relaxed. These vacua persist in the presence of non-perturbative effects. In all cases, the vacua remain AdS.}


\maketitle

\section{Introduction}
The heterotic string compactifed on Calabi-Yau manifolds~\cite{Candelas:1985en} has long been a promising avenue towards realistic particle physics from string theory and, more recently, systematic model building tools have been developed and used to construct sizeable sets of heterotic standard models~\cite{Donagi:2004ia, Braun:2005bw, Braun:2005ux, Anderson:2011ns, Anderson:2012yf, Anderson:2013xka, Buchbinder:2014sya, Anderson:2014hia}. 

Moduli stabilisation and supersymmetry breaking in the heterotic string has been somewhat more problematic, particularly in comparison with IIB string theory. One reason is the absence of RR fluxes in the heterotic string which implies less flexibility for stabilising moduli through flux. Recently, it has been shown that the heterotic $E_8\times E_8$ gauge flux can be used to fill this gap and stabilise many, possibly all, of the complex structure moduli and all but one of the dilation and the K\"ahler moduli~\cite{Anderson:2010mh, Anderson:2011ty, Anderson:2011cza, Klaput:2012vv, Klaput:2013nla, Anderson:2014xha, delaOssa:2014cia, delaOssa:2014msa}. Nevertheless, heterotic (as indeed IIB) moduli stabilization has, traditionally, required the inclusion of non-perturbative effects, either from string instantons or gaugino condensation. This appears to remain true even when $E_8\times E_8$ gauge flux is used in the context of heterotic Calabi-Yau compactifications.

In the present paper, we are taking a different approach to heterotic moduli stabilisation by focusing on perturbative stabilisation of all geometric moduli. We do not explicitly discuss bundle moduli effects here but will comment on their stabilisation later. Our discussion will be in the context of the heterotic string on certain manifolds with $SU(3)$ structure, more specifically half-flat mirror manifolds~\cite{Gurrieri:2002wz, Gurrieri:2004dt, Gurrieri:2007jg}. Mirror symmetry suggest the existence of a large class of such manifolds although relatively few examples, including a number of homogeneous coset spaces, are known explicitly~\cite{Lust:1985be, Lust:1986ix, butruille, chatzistavrakidis2009reducing, Klaput:2011mz, Chatzistavrakidis:2012qb, Klaput:2012vv, Haupt:2014ufa}.

For such compactifications, the geometric flux leads to a perturbative superpotential for the $T$-moduli. We show that, for this superpotential, supersymmetric vacua do not exist at a finite value of the dilaton. Further, by adding D-terms associated to anomalous $U(1)$ symmetries, we find non-supersymmetric AdS vacua, with all $T$-moduli and the dilation stabilised in a purely perturbative fashion. (The analogue of complex structure moduli are absent in at least some of the models but, if present, can be stabilised by additional NS flux.) For appropriate parameter choices, these vacua are consistent in the sense of leading to weak coupling, large volume and a relatively small negative cosmological constant. Remarkably, for a certain structure of the D-terms which we refer to as ``aligned", these vacua can be found analytically. Specifically, the condition on the D-terms is given by equation \eqref{aligned}, which has the property that both the F-term and D-term potentials are extremised separately for our analytic extrema. For more general, non-aligned D-terms we can show numerically, by deformations starting from the aligned case, that these non-supersymmetric AdS vacua persist. Related work on non-supersymmetric string vacua can, for example, be found in Refs.~\cite{Gallego:2008sv, Chatzistavrakidis:2012qb, Blaszczyk:2014qoa, Kallosh:2014oja, Angelantonj:2014dia}.

The vacua we find in this way are stable in the conventional sense, that is, they have a positive definite mass matrix rather than merely satisfying the Breitenlohner-Freedman condition~\cite{Breitenlohner1982197}. It is, therefore, conceivable that they can be lifted to stable dS vacua. We have investigated possible uplift mechanisms, namely the deformation of the D-terms away from alignment and the addition of perturbative and non-perturbative effect. Unfortunately, we have not been able to find dS vacua with any of these methods. 

Our vacua break supersymmetry perturbatively and, given the absence of small exponential factors as would arise from non-perturbative effects, the scale of this breaking is high and close to the string scale. Hence, models based on this type of breaking do not have low-energy supersymmetry and, at present, it is not clear to us how the hierarchy problem might be addressed in a meaningful way. However, such models with a high scale of supersymmetry breaking may come to be seen as a viable option if low-energy supersymmetry is not found experimentally.  

The paper is organized as follows. In Section \ref{sec:model}, we begin by describing the general class of models we are considering, laying out some technical details and discussing an illustrative example as we go along. Section \ref{sec:analytic} presents the general analytic {\it non-supersymmetric} AdS vacuum for these models. An explicit example is presented in Section~\ref{sec:exalligned}. The possibilities of lifting these solutions to dS space are discussed in Section \ref{sec:dSsearch} and we conclude in Section \ref{sec:concl}.

\section{The model}
\label{sec:model}
Let us now introduce the relevant class of $N=1$ supergravity models.  These models arise as the low energy effective theory of heterotic compactifications on half-flat mirror manifolds of the kind studied, for example, in Refs.~\cite{Lust:1985be, Lust:1986ix, Gurrieri:2004dt, micu2004heterotic, Gurrieri:2007jg, KashaniPoor:2007tr, zoupanos2, chatzistavrakidis2009reducing, Chatzistavrakidis:2009mh, Lukas:2010mf, Klaput:2011mz, CyrilThesis, Klaput:2012vv}. For simplicity, we will ignore the matter sector of this theory, and focus on the moduli, or gravitational sector. We also assume that the analogue of complex structure moduli have already been fixed either by bundle effects or NS flux or are absent as for half-flat coset models. We will, therefore, be focusing on the $T$-moduli and the dilaton $S$.

\subsection{Basic set-up}
As mentioned above, the relevant field content consists of the $T$-moduli $T^i$ and the dilation $S$ which reside in chiral multiplets and whose scalar components are broken up into real and imaginary parts as
\begin{equation} 
T^i=t^i+i\tau^i\;,\quad S=s+i\sigma\; ,
\end{equation}
where $i,j,\dots=1,\ldots ,n$. Here, $\tau^i$ and $\sigma$ are axions and $t^i$ are the geometric moduli. It is notationally useful to combine these fields into a single entity by defining $T^0=S$, $t^0=s$, $\tau^0=\sigma$ and by writing
\begin{equation}
T^I=t^I+i\tau^I\; ,
\end{equation}
where $I,J,\dots=0,1,\ldots, n$. In this way, for much of our general set-up, the dilaton and the $T$-moduli can be treated on the same footing. 

In this language, the K\"ahler potential~\cite{Gurrieri:2004dt}  can be written as
\begin{equation}
 K=-\ln\kappa\; ,\quad  \kappa = d_{IJKL}t^It^Jt^Kt^L\; , \label{Kpotdef}
\end{equation}  
where $\kappa$ is a quartic pre-potential and $d_{IJKL}$ are numbers whose only non-zero components are $d_{0ijk}=d_{ijk}$ and symmetric permutations thereof. The numbers $d_{ijk}$ are determined by the underlying half-flat manifold and can be thought of as the analogue of intersection numbers. We use the standard notation 
$\kappa_I=d_{IJKL}t^Jt^Kt^L$, $\kappa_{IJ}=d_{IJKL}t^Kt^L$ etc. for the derivatives of the pre-potential as well as $K_I=\partial K/\partial T^I$, $K_{IJ}=\partial^2 K/\partial T^I\partial T^J$ etc. for the derivatives of the K\"ahler potential. It is also useful to introduce the ``lower-index" fields
\begin{equation*}
 t_I\equiv K_{IJ}t^J=\frac{\kappa_I}{\kappa}\; ,
\end{equation*}
which satisfy $t^It_I=1$. Further, it follows that
\begin{equation}
 K_I=-\frac{2\kappa_I}{\kappa}\; ,\quad K_{IJ}=-3\left(\frac{\kappa_{IJ}}{\kappa}-\frac{4\kappa_I\kappa_J}{3\kappa^2}\right)\; .
\end{equation}
The superpotential for this class of models~\cite{Gurrieri:2004dt,Klaput:2012vv}  is given by
\begin{equation}
 W=w+e_IT^I\; , \label{eq:W}
\end{equation}
where $w$ and $e_I$ are real constants. In fact, $e_0=0$ since the superpotential needs to be dilaton-independent at the perturbative level. The remaining numbers $e_i$ encode the geometric flux of the underlying half-flat manifold. The constant $w$ can result either from harmonic NS flux (in which case it should eventually be thought of as complex-structure dependent) or $\alpha'$ corrections induced by the heterotic Bianchi identity~\cite{Klaput:2012vv}. 

In addition, we assume the existence of anomalous $U(1)$ symmetries (which can originate from internal line bundles) under which the moduli transform non-linearly as
\begin{equation*}
  \delta T^I=i\epsilon^ac^{I}_a\; ,
\end{equation*}
where $\epsilon^a$ are the $U(1)$ group parameteres with the index $a=1,\ldots ,m$ labeling the various $U(1)$ symmetries and $c^{I}_a$ are constants. For the superpotential~\eqref{eq:W} to be invariant under those symmetries we require that
\begin{equation}
 e_Ic^{I}_a=0\label{inv}
\end{equation}
for all $a$. The associated D-terms (assuming the absence of $U(1)$ charged matter fields for simplicity) are given by
\begin{equation}
 D_a=-c^{I}_aK_I=2c^{I}_at_I\; . \label{Dterm}
\end{equation} 
Finally, we require the gauge kinetic function which we write as
\begin{equation}
f_{ab}=\beta_IT^I\delta_{ab}+\kappa_{IJK}\gamma^Ic_a^Jc_b^k\; , \label{f}
\end{equation}
where $(\beta_I)=(1,\beta_i)$ and $\gamma^0$ is the only non-zero component of $\gamma^I$. For practical calculations later on we will neglect the corrections to this gauge kinetic function and simply use the leading expression $f_{ab}=S\delta_{ab}$. For consistent heterotic compactifications the validity of the strong coupling expansion~\cite{Witten:1996mz} is required which implies that $t^i\ll s$.  In this case, the threshold correction to the gauge kinetic function~\eqref{f} are small and can indeed be neglected.  

\subsection{The scalar potential}
The F-terms for the above class of models are given by
\begin{equation}
 F_I=W_I+K_IW=e_I-2t_IW\; .
\end{equation}
The F-terms equations, $F_I=0$, can easily be solved and, provided $w\neq 0$, result in
\begin{equation}
 t_I=-\frac{1}{2w}e_I\; ,\quad e_I\tau^I=0\; , \label{susymin}
\end{equation}
with $W=-w$ at this solution. Note, by inserting into Eq.~\eqref{Dterm} and using Eq.~\eqref{inv}, that the D-term equations, $D_a=0$, are automatically satisfied for this solution as it should be the case. In principle, this is a perfectly good supersymmetric anti-de Sitter vacuum. However, since we have set $e_0=0$ in order to have a dilaton-independent superpotential it follows that $t_0=1/(4s)=0$ and, hence, that we are at zero coupling. We conclude that the above class of models does not have supersymmetric vacua for finite field values\footnote{In Ref.~\cite{Klaput:2012vv}  a non-perturbative potential from gaugino condensation has been added to the superpotential \eqref{eq:W}. In this case, it is possible to obtain a supersymmetric AdS vacuum. As the purpose of this paper is to study {\it perturbative} solutions, we shall not pursue this route here.} .
 
For the scalar potential we find
\begin{eqnarray}
 V&=&V_F+V_D\\
 V_F&=&\frac{1}{\kappa}\left[K^{IJ}e_Ie_J-3(e_It^I)^2+w^2+2we_It^I+(e_I\tau^I)^2\right]\\
 V_D&=&4\sum_{a,b}f_R^{ab}c_a^Ic_b^Jt_It_J\; .
\end{eqnarray} 
It is relatively easy to discuss the fate of the axions, $\tau^I$. Provided that at least one of the geometric fluxes $e_I$ is non-zero, which we assume, the above scalar potential is minimized in the axion directions iff
\begin{equation*}
 e_I\tau^I=0\; ,
\end{equation*}
and, in this case, only the axion combination $e_I\tau^I$ is stabilized while the other axions remain flat directions. Some of these axions will be ``eaten" by the $U(1)$ gauge bosons which are massive. To see how this works consider the mass matrix
\begin{equation*}
 M_{ab}=c_a^Ic_b^JK_{IJ}
\end{equation*}
for these gauge bosons which follows from the kinetic terms of the $\tau^I$. If the rank of this matrix is maximal all $U(1)$ gauge bosons are massive. In general, however, this does not need to be the case. Since the K\"ahler metric is positive definite, we have ${\rm rk}(M)={\rm rk}(C)$ heavy $U(1)$ gauge bosons, where $C$ is the matrix $C=(c_a^I)$. Hence, we start with $n+1$ axions $\tau^I$, one of which is stabilized, ${\rm rk}(C)$ will are absorbed by the $U(1)$ gauge bosons and
\begin{equation*}
 n-{\rm rk}(C)
\end{equation*}
 remain as flat directions. This number may be zero but even if it is not there is no real problem for moduli stabilization since axions have a compact field space and will, almost inevitably, be stabilized. 

\subsection{A simple example}
In order to get a feel for the models, let us consider a very simple example with three fields, $S=s+i\sigma$, $T=t+i\tau$ and $U=u+i\nu$, a pre-potential\footnote{The pre-potential we consider here is significantly simpler than the ones usually arising from compactifications, even for simple coset constructions \cite{Klaput:2012vv}, but it is sufficient for the points we wish to make.}
\begin{equation}
 \kappa=st^2u\; , \label{kappaex}
\end{equation} 
a superpotential taking the assumed form coming from string-compactifications on torsional half-flat manifolds
\begin{equation*}
 W=w+eT\; ,
\end{equation*}
and a single $U(1)$ symmetry under which $T$ is invariant and $S$ and $U$ transform. The D-term has the structure
\begin{equation*}
 D=\frac{c}{u}+\frac{b}{s}\; ,
\end{equation*}
with real constants $c$ and $b$. For the scalar potential we find
\begin{equation*}
 V=\frac{1}{st^2u}\left[w^2-2ewt-e^2t^2+e^2\tau^2\right]+\frac{1}{s}\left[\frac{c}{u}+\frac{b}{s}\right]^2\; ,
\end{equation*}
where, for simplicity, we have only considered the leading term, $f=S$, of the gauge kinetic function in the D-term part of the potential. A quick algebraic calculation using the Stringvacua package~\cite{Gray:2008zs} shows that this potential has three stationary points. Two of these arise at unphysical field values and the third is given by
\begin{equation}
\label{eq:solNonSUSY}
 s=\frac{4bc}{e^2}\; ,\quad t=\frac{w}{e}\; ,\quad u=\frac{4c^2}{3e^2}\; ,\quad \tau=0\; .
\end{equation} 
The two other axions, $\sigma$ and $\nu$, remain flat directions but one combination of these fields is absorbed by the $U(1)$ vector boson. In order for all field values to be positive we have to require that $bc>0$ and $we>0$. Furthermore, all field values should be large compared to one so that the model is at weak coupling and the supergravity approximation is valid. This can clearly be achieved for suitable choices of the parameters. 

We define the two constants
\begin{equation*}
 k=\frac{3e^8}{8bc^3w^2}\; ,\quad r=\frac{3e^2w^2}{64bc^3}\; ,
\end{equation*}
which are both positive provided the aforementioned condition $bc>0$ is satisfied. In terms of these constants, the Hessian can be written as
\begin{equation*}
 H=k\left(\begin{array}{llllll}\frac{cr}{b}&0&r&0&0&0\\0&1&0&0&0&0\\r&0&\frac{9br}{c}&0&0&0\\0&0&0&0&0&0\\0&0&0&0&1&0\\0&0&0&0&0&0\end{array}\right)\; ,
\end{equation*}  
where the ordering of fields is $(s,t,u,\sigma,\tau,\nu)$. Obviously, there are two zero eigenvalues, corresponding to the axions $\sigma$ and $\tau$, two eigenvalues with size $k$ and two further eigenvalues
\begin{equation*}
\frac{kr}{2bc}\left(9b^2+c^2\pm\sqrt{81b^4-14b^2c^2+c^4}\right)\; .
\end{equation*}
For $bc>0$ these eigenvalues are positive so we have a minimum. The cosmological constant
\begin{equation*}
 V_{\rm min}=-\frac{e^6}{8bc^3}\; .
\end{equation*} 
is always negative while the F-terms at the minimum are given by
\begin{equation*}
 (F_S,F_T,F_U)=-e\left(\frac{ew}{4bc},1,\frac{3ew}{4c^2}\right)=-e\left(\frac{t}{s},1,\frac{u}{t}\right)\:,
\end{equation*} 
so that supersymmetry is broken. Hence, we have found a perturbative, non supersymmetric AdS vacuum. As the above expression shows, $F_S$ and $F_U$ can be made small by a suitable hierarchy of values for the moduli. However, $F_T=-e$ is more problematic. From a supergravity point of view the parameter $e$ can, of course, chosen to be small. However, in a string context, $e$ corresponds to the intrinsic torsion of the compactification manifold and is quantised. Therefore, separation of scales between the fundamental and the supersymmetry breaking scales is difficult to achieve in this model. Our general analysis later on shows that this is not a general property of our class but that separation of scales is possible in some cases. 

\section{General analytic minimum}
\label{sec:analytic}
Does the vacuum for the simple three-field model we have just found represent a special case which arises for a small number of fields or does it indicate a property of the entire class of models? Commutative algebra and numerical methods quickly run into the ground for a larger number of fields, so in order to answer this question we should search for a general analytical solution. 

A starting point is suggested by the structure of the supersymmetric vacuum~\eqref{susymin} which led to lower-index fields $t_I$ being proportional to the geometric flux parameters $e_I$. This failed to provide a vacuum for a finite dilaton value since $e_0=0$ implies $t_0=1/(4s)=0$. The obvious course of action is to slightly weaken this Ansatz and demand that only the $T$-moduli satisfy
\begin{equation}
 t_i=\alpha e_i\;, \label{ans}
\end{equation}
where $\alpha$ is a constant to be determined, while the dilation $s$ remains arbitrary for now. 

\subsection{Some useful very special geometry relations}
Before we explore the implications of this Ansatz it is useful to collect a few very special geometry results for the cubic pre-potential ${\cal K}$ which is defined by
\begin{equation}
 \kappa=4s{\cal K}\;,\quad {\cal K}=d_{ijk}t^it^jt^k\; ,
\end{equation} 
where we recall that $\kappa$ is the quartic pre-potential introduced in Eq.~\eqref{Kpotdef}. As usual, we denote the derivatives of ${\cal K}$ by
${\cal K}_i=d_{ijk}t^jt^k$, ${\cal K}_{ij}=d_{ijk}t^k$ and ${\cal K}_{ijk}=d_{ijk}$. With this notation, the first derivative of the K\"ahler potential $K=-\ln \kappa$ and the K\"ahler metric can be written as
\begin{equation}
\begin{array}{lllllllllll}
K_0&=&-\frac{1}{2s}&\quad\quad& K_i&=&-\frac{3{\cal K}_i}{2{\cal K}}&\quad\quad&&&\\
K_{00}&=&\frac{1}{4s^2}&\quad\quad&K_{0i}&=&0&\quad\quad& K_{ij}&=&-\frac{3}{2}\left(\frac{{\cal K}_{ij}}{{\cal K}}-\frac{3}{2}\frac{{\cal K}_i{\cal K}_j}{{\cal K}^2}\right)\; .
\end{array}
\end{equation}
The lower-index fields $t_I=K_{IJ}t^J$, explicitly given by
\begin{equation}
 t_0=K_{00}t^0=\frac{1}{4s}\;,\qquad t_i=K_{ij}t^i=\frac{3{\cal K}_i}{4{\cal K}}\; ,
\end{equation} 
then satisfy the useful relations
\begin{equation}
 t_it^i=\frac{3}{4}\:,\qquad  \frac{\partial t_i}{\partial t^j}=-K_{ij}\;.
\end{equation} 
It will also be convenient to introduce the notation
\begin{equation}
 K_{i_1\dots i_p}=\frac{\partial}{\partial T^{i_1}}\dots \frac{\partial}{\partial T^{i_p}}K
                          =\frac{1}{2^p}\frac{\partial}{\partial t^{i_1}}\dots \frac{\partial}{\partial t^{i_p}}K
\end{equation}
for the $p^{\rm th}$ derivatives of the K\"ahler potential. These tensors are obviously completely symmetric and indices will be lowered and raised by the K\"ahler metric $K_{ij}$ and its inverse. These tensors satisfy a number of relations which include
\begin{equation}
 K_it^i=-\frac{3}{2}\,,\;\; K_{ij}t^j=-\frac{1}{2}K_i\,,\;\; K_{ijk}t^k=-K_{ij}\,,\;\; K_{ijkl}t^l=-\frac{3}{2}K_{ijk}\; .
\end{equation} 

\subsection{The scalar potential}
We now come back to our supergravity theory and recall that the superpotential is given by
\begin{equation}
 W=w+e_it^i
\end{equation}
For the F- and D-terms we find
\begin{equation}
 F_0=K_0W=-\frac{1}{2s}W\;,\qquad F_i=W_i+K_iW=e_i -2t_iW\;,\qquad D_a=2c_a^it_i+\frac{c_a^0}{2s}\; ,
\end{equation} 
while the scalar potential is given by
\begin{eqnarray}
 V&=&V_F+V_D\\
 V_F&=&\frac{1}{4s{\cal K}}\left[K^{kl}e_ke_l-3(e_kt^k)^2+w^2-2we_kt^k\right]\\
 V_D&=&\frac{4}{s}\sum_a\left(\frac{c^{a0}}{4s}+c^{ak}t_k\right)^2\; .
\end{eqnarray}
Here we have already set $e_i\tau^i=0$ for the minimum in the axion directions and, for simplicity, we have also used the lowest order gauge kinetic function $f=S$ in the D-term potential. For the various derivatives of the scalar potential we find
\begin{eqnarray}
 \frac{\partial V_F}{\partial s}&=&-\frac{1}{s}V_F\\
 \frac{\partial V_D}{\partial s}&=&-\frac{4}{s^2}\sum_a\left(\frac{c_a^{0}}{4s}+c_a^{k}t_k\right)^2-\frac{2}{s^3}\sum_ac^{a0}\left(\frac{c_a^{0}}{4s}+c_a^{k}t_k\right)\\
 \frac{\partial V_F}{\partial t^i}&=&\frac{1}{4s{\cal K}}\left[-2\left(K_i^{kl}+2t_iK^{kl}\right)e_ke_l+4\left(3(e_kt^k)^2-w^2+2we_kt^k\right)t_i\right.\nonumber\\
 &&\left.\qquad\;\,-6(e_kt^k)e_i-2we_i\right]\\
 \frac{\partial V_D}{\partial t^i}&=&-\frac{8}{s}K_{ik}\sum_ac_a^{k}\left(\frac{c_a^{0}}{4s}+c_a^{l}t_l\right)
\end{eqnarray} 

\subsection{The vacua}
After this preparation, we are now ready to come back to the Ansatz~\eqref{ans} which we write as
\begin{equation}
 t_i=\frac{3{\cal K}_i}{4{\cal K}}=\alpha e_i\; , \label{ansatz}
\end{equation}
where $\alpha$ is a real constant to be fixed later. Note, that Eq.~\eqref{ansatz} represents a set of, generally complicated, algebraic equations for the actual fields $t^i$ with upper indices which can be explicitly solved for any given model once the ``intersection numbers" $d_{ijk}$ are known. The dilaton value will be determined later. Let us first evaluate the potential and its first derivates for the Ansatz~\eqref{ansatz}. All quantities evaluated for the field values fixed by Eq.~\eqref{ansatz} are denoted by a subscript $0$. We find
\begin{eqnarray}
 V_F|_0&=&-\frac{v}{s}\;,\quad v=\frac{1}{4{\cal K}_0}\left(\frac{15}{16\alpha^2}+\frac{3w}{2\alpha}-w^2\right)\label{VF0}\\
 V_D|_0&=&\frac{c^2}{4s^3}\;,\quad c^2=\sum_a\left(c^{a0}\right)^2\label{VD0}\\
 \left.\frac{\partial V_F}{\partial s}\right|_0&=&\frac{v}{s^2}\\
 \left.\frac{\partial V_D}{\partial s}\right|_0&=&-\frac{3c^2}{4s^4}\\
 \left.\frac{\partial V_F}{\partial t^i}\right|_0&=&\frac{1}{4s{\cal K}_0}\left(\frac{5}{4\alpha}+4w-4w^2\alpha\right)e_i  \label{VFt}\\
 \left.\frac{\partial V_D}{\partial t^i}\right|_0&=& -\frac{2}{s^2}K_{ik}|_0\sum_ac_a^{0}c_a^{k} \label{VDt}
\end{eqnarray} 
There are two features of this result which are remarkable. First, the derivatives of the D-term simplify considerably by virtue of the condition $c^{ak}t_k=\alpha c^{ak}e_k=0$ which follows from gauge invariance of the superpotential. Secondly, for our Ansatz the derivatives $\partial V_F/\partial t^i$ become proportional to the geometric fluxes $e_i$ and, therefore, effectively reduce to one equation. One obvious problem is that the derivatives $\partial V_D/\partial t^i$ of the D-term potential have a different, more complicated structure and, in particular, are not proportional to $e_i$. However, as we will now show, for our Ansatz to work, the $t^i$ derivatives of $V_F$ and $V_D$ have to vanish independently. To see this contract both Eq.~\eqref{VFt} and \eqref{VDt} with $t^i=\alpha K^{ij}e_j$. From the gauge invariance condition~\eqref{inv} it follows that 
\begin{equation}
 t^i\frac{\partial V_D}{\partial t^i}=-\frac{2}{s}e_k\sum_ac_a^0c_a^k=0\; ,
\end{equation} 
and, therefore, for a stationary point of $V$ we require that
\begin{equation}
 t^i\frac{\partial V_F}{\partial t^i}=\frac{1}{4s{\cal K}_0}\left(\frac{5}{4\alpha}+4w-4w^2\alpha\right)e_iK^{il}|_0e_l\stackrel{!}{=}0\; .
\end{equation}
Since the K\"ahler metric $K_{ij}$ is positive definite this can only be satisfied provided
\begin{equation}
\frac{5}{4\alpha}+4w-4w^2\alpha=0\:. \label{alphaeq}
\end{equation}
which, in turn, implies that $\partial V_F/\partial t^i|_0=0$. A stationary point of the potential then requires that $\partial V_D/\partial t^i|_0\stackrel{!}{=}0$ and this can only be achieved provided  the D-terms satisfy the condition
\begin{equation}
 \sum_ac_a^{0}c_a^{k}=0\; . \label{aligned}
\end{equation}
We will refer to D-terms satisfying this equation as ``aligned" and for now we assume this property.

Then, there is a stationary point of $V$ with
\begin{equation}
 t_i=\alpha e_i\;,\quad s^2=\frac{3c^2}{4v}\; ,\quad e_i\tau^i=0\; , \label{tssol}
\end{equation}
where $v$ and $c^2$ have been defined in Eqs.~\eqref{VF0}, \eqref{VD0} and Eq.~\eqref{alphaeq} leads to the two $\alpha$ values
\begin{equation}
\alpha= \left\{\begin{array}{rrr}-\frac{1}{4w}&\quad &\mbox{case 1}\\\frac{5}{4w}&\quad&\mbox{case 2}\end{array}\right.\quad\Rightarrow\quad
v= \left\{\begin{array}{rrr}\frac{2w^2}{{\cal K}_0}&\quad &\mbox{case 1}\\\frac{w^2}{5{\cal K}_0}&\quad&\mbox{case 2}\end{array}\right. \label{alphaval}
\end{equation}
Here, ${\cal K}_0$ is the cubic pre-potential evaluated on the above solution for the moduli $t^i$.  Of course, there are some restrictions on the moduli $t^i$ which must reside in the ``K\"ahler" cone of the underlying manifold. The low-energy test for this is that ${\cal K}_0>0$ (so that the internal volume is positive) and that the K\"ahler metric $K_{ij}|_0$ is positive definite (so that the kinetic terms are well-defined). The fluxes $e_i$ and $w$ have to be chosen such that this is indeed the case, but there is no general reason why this should not be possible. However, it is clear that the moduli $t^i$ can only be in the K\"ahler cone for at most one of the two cases in Eq.~\eqref{alphaval}, depending on the signs of $w$ and $e_i$. We will discuss an explicit example later on and show that there is no general obstruction to a consistent choice of parameters. Provided such a choice has been made, both values for $v$ are positive and, hence, the dilaton equation in \eqref{tssol} leads to a physically sensible value for the dilaton. 
The potential value at this solution
\begin{equation}
\label{potAnal}
 V|_0=-\frac{2v}{3s}<0
\end{equation}
is always negative.

We should also discuss if weak coupling and large radii can be achieved by suitable parameter choices. If we denote a typical geometric flux by $e$ and a typical $t^i$ modulus by $t$ then we have the following rough scaling relations:
\begin{equation}
 s^2\sim \frac{c^2w}{e^3}\;,\quad t\sim \frac{w}{e}\;,\quad V|_0\sim -\frac{w^2}{st^3}\; .
\end{equation}
This shows that weak coupling and large volume can indeed be arranged, the latter of course being essential for the validity of the supergravity approximation and that we can achieve $t\ll s$ so that the strong-coupling expansion is valid. In this case, since the threshold corrections to the gauge kinetic function are negligible, the gauge couplings are proportional to $1/s$ and are, hence, in the perturbative regime. Moreover, provided $s$ and $t$ are large the absolute value of the cosmological constant is suppressed and somewhat below the fundamental scale.

\subsection{Stability}
Crucially, of course, stability has to be checked for this solution and this leads to a somewhat tedious calculation of the second derivatives of $V$. However, this calculation can be much simplified by using the very special geometry relations listed previously. We find
\begin{equation}
\begin{array}{lllllll}
 \left.\frac{\partial^2 V_F}{\partial t^i\partial t^j}\right|_0&=&\frac{1}{4s{\cal K}_0}\left(AK_{ij}+Bt_it_j\right)&\quad&
 \left.\frac{\partial^2 V_D}{\partial t^i\partial t^j}\right|_0&=&\frac{8}{s}K_{ik}|_0K_{jl}|_0\sum_ac^{ak}c^{al}\\
 \left.\frac{\partial^2 V_F}{\partial s^2}\right|_0&=&-\frac{2v}{s^3}&\quad&
 \left.\frac{\partial^2 V_D}{\partial s^2}\right|_0&=&\frac{3c^2}{s^5}\\
  \left.\frac{\partial^2 V_F}{\partial s\partial t^j}\right|_0&=&0&\quad&\left.\frac{\partial^2 V_D}{\partial s\partial t^j}\right|_0&=&0
 \end{array}
\end{equation} 
where
\begin{equation}
 A=-\frac{7}{4\alpha^2}+4w^2-\frac{6w}{\alpha}\;,\quad B=16w^2-\frac{8w}{\alpha}-\frac{1}{\alpha^2}\; . \label{ABdef}
\end{equation}
First of all, we note that the Hesse matrix does not mix the dilaton and the $t$ moduli. Inserting the dilaton solution from Eq.~\eqref{tssol} we find
\begin{equation}
 \left.\frac{\partial^2 V}{\partial s^2}\right|_0=\frac{3c^2}{2s^5}\; .
\end{equation}
so that the dilation is stable. Combining the above results, and introducing the notation $C=(c^{ak})$, $G=(K_{ij})$, ${\bf t}=(t^i)$ and ${\bf s}=(t_i)$ we find in the $t$ directions
\begin{equation}
 {\bf v}^T\left.\frac{\partial^2 V}{\partial{\bf t}^2}\right|_0{\bf v}=\frac{1}{s}\left[\frac{1}{4{\cal K}_0}\left(A {\bf v}^TG{\bf v}+B|{\bf s}^T{\bf v}|^2\right)+8|CG{\bf v}|^2\right]\; , \label{Hesse}
\end{equation}
for any vector ${\bf v}=(v^i)$. Note that the last term is positive and the signs of the other two terms are given by the signs of $A$ and $B$, respectively. Inserting the solutions~\eqref{alphaval} for $\alpha$ into the definition~\eqref{ABdef} of $A$ and $B$ we find
\begin{equation}
\alpha= \left\{\begin{array}{rrr}-\frac{1}{4w}&\quad &\mbox{case 1}\\\frac{5}{4w}&\quad&\mbox{case 2}\end{array}\right.\quad\Rightarrow\quad
(A,B)=\left\{\begin{array}{ccr}(0,32w^2)&\quad &\mbox{case 1}\\\left(-\frac{48w^2}{25},\frac{224w^2}{25}\right)&\quad&\mbox{case 2}\end{array}\right.
\end{equation}
Hence, in case 1 the Hesse matrix is always positive definite and we have a stable AdS vacuum. Case 2 this is not quite so straightforward since $A$ is negative. However, the numerical values are such that it seems likely the second and third term in~\eqref{Hesse} which are positive will overcome the negative contribution of the first term. Also, the last term in \eqref{Hesse} depends on $C=(c^{ak})$ which does not enter the solution anywhere else, so by increasing its value it should be possible to obtain a positive definite Hesse matrix. To be sure, this has to be explicitly checked and we will do this for our simple example below which indeed leads to a stable vacuum for both cases.

Finally, we should discuss supersymmetry breaking for our vacua. For our vacua we find
\begin{eqnarray}
 D_a&=&\frac{c^{a0}}{2s}\;,\quad\\
 F_0&=&-\frac{1}{2s}\left(w+\frac{3}{4\alpha}\right)=\left\{\begin{array}{cll}\frac{w}{s}&&\mbox{case 1}\\-\frac{4w}{5s}&&\mbox{case 2}\end{array}\right.\\
 F_i&=&-\frac{1}{2}(1+4\alpha w)e_i=\left\{\begin{array}{cll}0&&\mbox{case 1}\\-3e_i&&\mbox{case 2}\end{array}\right.\; .
\end{eqnarray}
Evidently, these patterns are very different for the two cases. In case 1, F-term supersymmetry breaking arises only in the dilaton direction and all non-zero terms are proportional to $1/s$. Hence, in this case the supersymmetry breaking scale can be below the fundamental scale, a consistency condition for starting with a supersymmetric field theory in the first place. It should however be noted that for typical physical values of $s$ this separation is only by an order of magnitude or so. In case 2, on the other hand, all F-terms are non-zero and the F-terms in the $T$-directions are proportional to $e_i$. In a string theory context, these quantities are quantised and a supersymmetry breaking scale below the fundamental scale seems difficult to achieve. \\[3mm]
In summary, we have found generic, non-supersymmetric AdS vacua at weak coupling and sufficiently large volume for our class of models. These arise in two cases, depending on the signs of the superpotential parameters. In the first case, these vacua are guaranteed to be minima and the supersymmetry breaking scale and the fundamental scale can be separated. In the second case, it is likely that minima can be achieved for suitable parameter choices but scale separation seems difficult to realise. 

\section{An example with aligned D-terms}
\label{sec:exalligned}
For illustration and to show that all constraints can indeed be satisfied let us discuss a very simple model with aligned D-terms. Consider the simple field content $(T^I)=(S,T,U)$ with $S=s+\sigma$, $T=t+i\tau$ and $U=u+i\nu$ and a K\"ahler potential
\begin{equation*}
 {\cal K}=t^2u\; ,
\end{equation*}
as before. This means the lower index fields are given by
\begin{equation*}
 t_0=\frac{1}{4s}\;,\quad t_1=\frac{1}{2t}\;,\quad t_2=\frac{1}{4u}\; .
\end{equation*} 
We use a flux vector ${\bf e}=(0,e,\epsilon)$ so that the superpotential reads
\begin{equation*}
 W=w+eT+\epsilon U\; .
\end{equation*}  
Further, we introduce two D-terms with ${\bf c}_1=2(b,c/e,-c/\epsilon)$ and ${\bf c}_2=2(-b,c/e,-c\epsilon)$ where the three entries refer to $(s,t,u)$. Note that indeed ${\bf c}_1\cdot{\bf e}={\bf c}_2\cdot{\bf e}=0$, as required for gauge invariance of the superpotential. In addition, we have
\begin{equation*}
 c_1^0c_1^k+c_2^0c_2^k=0
\end{equation*}
so that Eq.~\eqref{aligned} is satisfied and the two D-terms are aligned. From Eq.~\eqref{Dterm} the two D-terms read explicitly
\begin{equation*}
 D_1=\frac{2c}{et}-\frac{c}{\epsilon u}+\frac{b}{s}\; ,\quad D_2=\frac{2c}{et}-\frac{c}{\epsilon u}-\frac{b}{s}\; .
\end{equation*}
The scalar potential $V=V_F+V_D$ is then given by
\begin{equation}
 V_F=\frac{1}{st^2u}\left(\epsilon^2u^2-e^2t^2-6e\epsilon tu-2w\epsilon u-2wet+w^2\right)\; ,\quad
 V_D=\frac{1}{s}\left(D_1^2+D_2^2\right)\; , \label{potex}
\end{equation} 
where the axions have already been removed by integrating out the massive axion direction $e\tau+\epsilon \nu=0$. The two other axions are absorbed by the gauge bosons of the two $U(1)$ symmetries so for this example there are no massless axion directions left over.\\[3mm] 
From the general expression Eq.~\eqref{tssol}, the solution is given by
\begin{equation*}
 s^2=\frac{3b^2}{2v}\;,\quad t=\frac{1}{2e\alpha}\;,\quad u=\frac{1}{4\epsilon\alpha}\; ,
\end{equation*}
where $\alpha$ and $v$ have two values
\begin{equation}
\alpha=\left\{\begin{array}{cll}-\frac{1}{4w}&\quad&\mbox{case 1}\\\frac{5}{4w}&\quad&\mbox{case 2}\end{array}\right.\quad\Rightarrow\quad
v=\left\{\begin{array}{cll}-\frac{e^2\epsilon}{2w}&\quad&\mbox{case 1 for }\frac{\epsilon}{w}<0\\
    \frac{25e^2\epsilon}{4w}&\quad&\mbox{case 2 for }\frac{\epsilon}{w}>0\end{array}\right.\; .
\end{equation}  
Note the requirement on the sign of $\epsilon/w$ in order to ensure that $v$ is positive, so that the value of the dilaton is real. It can be checked by direct calculation that the derivatives of the potential~\eqref{potex} indeed vanish for those field values. The value of the volume at the minimum is given by
\begin{equation}
{\cal K}_0=\left\{\begin{array}{cll}-\frac{4w^3}{e^2\epsilon}&\quad&\mbox{case 1}\\
                  \frac{4w^3}{125e^2\epsilon}&\quad&\mbox{case 2}\end{array}\right.
\end{equation}
while the K\"ahler metric is given by
\begin{equation}
 G|_0=\left\{\begin{array}{cll}\frac{1}{4}{\rm diag}\left(-\frac{e^2\epsilon}{3b^2w},\frac{e^2}{2w^2},\frac{e^2}{w^2}\right)&\quad&\mbox{case 1}\\
\frac{25}{4}{\rm diag}\left(\frac{e^2\epsilon}{6b^2w},\frac{e^2}{2w^2},\frac{e^2}{w^2}\right)&\quad&\mbox{case 2}\end{array}\right.\; .
\end{equation} 
For the correct choice of the sign of $\epsilon/w$, as above, the volume is indeed positive and the K\"ahler metric is positive definite for both cases.
The Hesse matrix is somewhat more complicated but again turns out to be positive definite in both cases for the right sign of $\epsilon/w$.

\section{Searching for dS vacua}
\label{sec:dSsearch}
In this section, we would like to discuss generalisations and extensions and, in particular, address the problem of lifting the AdS vacua we have found to dS vacua. In principle, a number of possibilities come to mind. These include the deformation of the D-terms away from the aligned configuration, the addition of radiative corrections to the scalar potential and non-perturbative effects.\\[3mm]
It is useful to begin with a general discussion of the dilaton effective potential. After integrating out the $T$-moduli this potential has the general form
\begin{equation}
 V_{\rm eff}(s)=\frac{C_1}{s}+\frac{C_2}{s^2}+\frac{C_3}{s^3}\; ,
\end{equation}
where the first term results from the F-term potential and the second and third term from the D-terms. It is straightforward to show that a potential of this general form only has stable dS vacua if $C_1>0$, $C_2<0$ and $C_3>0$. If any of these conditions is violated the stationary points are either AdS or unstable. As can be seen from Eqs.~\eqref{VF0} and \eqref{VD0}, the analytic vacuum leads to $C_1=-v<0$ and $C_2=0$ which violates the above conditions for a dS vacuum. It seems difficult to change the sign of $C_1$, the coefficient of the term with the lowest suppression by inverse powers of $s$, using a small correction and this points to the main source of the problem.\\[3mm]
We begin with de-aligning the D-terms. This has to be done in the context of a specific case and we consider the example presented in the previous section. For this model, we modify the second D-term to 
\begin{equation}
 D_2=\frac{2c}{et}-\frac{c}{\epsilon u}-\frac{\delta b}{s}
\end{equation}
where $\delta$ is a real parameter and $\delta=1$ corresponds to the aligned case. All other features of the model are kept unchanged. Starting with the analytic vacuum available for $\delta=1$, we can now gradually change $\delta$ away from $1$ and minimise the potential numerically at each step. In this way we find that AdS minima exist for up to order one changes of $\delta$. This works starting from vacua for both case 1 and case 2. Hence, the existence of AdS minima is not an artefact of aligned D-terms but persists more generally, although analytic solutions are hard to find if the D-terms are not aligned. However, we have not been able to lift the AdS vacua to dS ones in this way - the cosmological constant remains negative when the D-terms are de-aligned.\\[3mm]  
Another option is to consider perturbative quantum corrections to the above theory. The one-loop corrections to a four-dimensional $N=1$ supergravity theory have been worked out in Refs.~\cite{srednicki1985, Gaillard:1993es, Gaillard:1996ms, Gaillard:1996hs}, and turn out to be rather complicated. Indeed, including the full one-loop correction will generate terms in the scalar potential up to $\OO(s^{-10})$. However, for phenomenological reasons we are interested in a vacuum with weak coupling, where $s\gg 1$. We therefore focus on the corrections to the $C_i$, where $i\leq 3$ assuming that the higher order corrections can be neglected. Unfortunately, the first order quantum corrections still appear to give the wrong sign for  $C_2$, and a search for dS vacua induced by the one-look corrected effective potential has so far been unsuccessful.\\[3mm]
Finally, we might consider non-perturbative corrections by modifying the superpotential to
\begin{equation}
W=w+e_I T^I+k\,e^{-p_IT^I}\; , \label{Wnonpert}
\end{equation}
while keeping all other features of the model unchanged. Of course, the combination $p_IT^I$ has to be invariant under the $U(1)$ symmetries, so $p_Ic_a^I=0$ for all $a$. It is worth mentioning that for a purely dilaton-dependent non-perturbative term, that is, $p_IT^I=pS$ (should this be gauge invariant) an analytic solution can still be found by the same Ansatz $t_i=\alpha e_i$, although the equations for $\alpha$ and the dilaton are now more complicated. A detailed analysis of this case shows that the AdS vacua persist and remains stable in the presence of the non-perturbative term but they cannot be lifted to dS vacua.

It may be worth discussing the possible fate of non-geometrical moduli such as bundle moduli. In this context, it is worth noting that the quantity $k$ in Eq.~\eqref{Wnonpert} is in general a function of complex structure and bundle moduli~\cite{Buchbinder:2002ic}. While the perturbative vacua for the $T^I$ are not destabilised by the presence of the non-perturbative terms, these contributions do, therefore, generate a potential for the bundle moduli. This potential may stabilise the bundle moduli but a detailed investigation of this issue is beyond the scope of the present paper.\\[3mm]
In summary, while the existence of AdS vacua is fairly robust it seems difficult to lift these to dS vacua. 

\section{Conclusion}
\label{sec:concl}
In this paper we have studied four-dimensional $N=1$ supergravity theories which arise in heterotic compactifications on {\it half-flat} manifolds. 
These models give rise to a superpotential linear in the K\"ahler moduli. In addition, anomalous $U(1)$ symmetries induced from internal split bundles can lead to dilaton and $T$-moduli dependent D-terms. For this class of models, plus an additional technical condition of ``aligned" D-terms, we have found analytic, supersymmetry breaking AdS minima. For suitable parameter choices, these minima are at weak coupling, at sufficiently large radii for the supergravity approximation to be valid and with the scale of supersymmetry breaking separated from the fundamental scale. Given that complex structure moduli may be absent in some cases or else be stabilised by NS flux these minima stabilise all geometric moduli of heterotic compactifications perturbatively. We have also verified, in the context of a specific example, that these minima are robust under deformations of the D-terms away from the aligned configuration. 

Since these minima are purely perturbative there are no exponential factors which might lead to a small supersymmetry breaking scale. Although the supersymmetry breaking scale can be suppressed, relative to the fundamental scale, it is not sufficiently small to be consistent with low-energy supersymmetry. Hence, for these models, supersymmetry is broken at a high scale and superpartner masses are far removed from the range which is currently accessible by experiment. However, they may well be of phenomenological interest if low-energy supersymmetry is not found. 

Unfortunately, we have not been able to find a way to lift to dS minima. Neither de-aligning the D-terms nor adding perturbative or non-perturbative corrections to the scalar potential leads to a positive cosmological constant. We leave this as a problem for future work.

\section*{Acknowledgements}
AL~is partially supported by the EPSRC network grant EP/l02784X/1 and by the STFC consolidated grant~ST/L000474/1. ZL is supported by the National Science Centre research grant DEC-2012/04/A/ST2/00099. EES~is supported by the ILP LABEX (under reference ANR-10-LABX-63), and by French state funds managed by the ANR within the Investissements dÕAvenir program under reference ANR-11-IDEX-0004-02. The authors would like to thank Michael Klaput for helpful discussions.

\bibliographystyle{JHEP}

\begin{thebibliography}{10}

\bibitem{Candelas:1985en}
P.~Candelas, G.~T. Horowitz, A.~Strominger, and E.~Witten, {\it {Vacuum
  Configurations for Superstrings}},  {\em Nucl.Phys.} {\bf B258} (1985)
  46--74.

\bibitem{Donagi:2004ia}
R.~Donagi, Y.-H. He, B.~A. Ovrut, and R.~Reinbacher, {\it {The Particle
  spectrum of heterotic compactifications}},  {\em JHEP} {\bf 0412} (2004) 054,
  [\href{http://xxx.lanl.gov/abs/hep-th/0405014}{{\tt hep-th/0405014}}].

\bibitem{Braun:2005bw}
V.~Braun, Y.-H. He, B.~A. Ovrut, and T.~Pantev, {\it {A Standard model from the
  E(8) x E(8) heterotic superstring}},  {\em JHEP} {\bf 0506} (2005) 039,
  [\href{http://xxx.lanl.gov/abs/hep-th/0502155}{{\tt hep-th/0502155}}].

\bibitem{Braun:2005ux}
V.~Braun, Y.-H. He, B.~A. Ovrut, and T.~Pantev, {\it {A Heterotic standard
  model}},  {\em Phys.Lett.} {\bf B618} (2005) 252--258,
  [\href{http://xxx.lanl.gov/abs/hep-th/0501070}{{\tt hep-th/0501070}}].

\bibitem{Anderson:2011ns}
L.~B. Anderson, J.~Gray, A.~Lukas, and E.~Palti, {\it {Two Hundred Heterotic
  Standard Models on Smooth Calabi-Yau Threefolds}},  {\em Phys.Rev.} {\bf D84}
  (2011) 106005, [\href{http://xxx.lanl.gov/abs/1106.4804}{{\tt
  arXiv:1106.4804}}].

\bibitem{Anderson:2012yf}
L.~B. Anderson, J.~Gray, A.~Lukas, and E.~Palti, {\it {Heterotic Line Bundle
  Standard Models}},  {\em JHEP} {\bf 1206} (2012) 113,
  [\href{http://xxx.lanl.gov/abs/1202.1757}{{\tt arXiv:1202.1757}}].

\bibitem{Anderson:2013xka}
L.~B. Anderson, A.~Constantin, J.~Gray, A.~Lukas, and E.~Palti, {\it {A
  Comprehensive Scan for Heterotic SU(5) GUT models}},  {\em JHEP} {\bf 1401}
  (2014) 047, [\href{http://xxx.lanl.gov/abs/1307.4787}{{\tt
  arXiv:1307.4787}}].

\bibitem{Buchbinder:2014sya}
E.~I. Buchbinder, A.~Constantin, and A.~Lukas, {\it {Non-generic Couplings in
  Supersymmetric Standard Models}},
  \href{http://xxx.lanl.gov/abs/1409.2412}{{\tt arXiv:1409.2412}}.

\bibitem{Anderson:2014hia}
L.~B. Anderson, A.~Constantin, S.-J. Lee, and A.~Lukas, {\it {Hypercharge Flux
  in Heterotic Compactifications}},
  \href{http://xxx.lanl.gov/abs/1411.0034}{{\tt arXiv:1411.0034}}.

\bibitem{Anderson:2010mh}
L.~B. Anderson, J.~Gray, A.~Lukas, and B.~Ovrut, {\it {Stabilizing the Complex
  Structure in Heterotic Calabi-Yau Vacua}},  {\em JHEP} {\bf 1102} (2011) 088,
  [\href{http://xxx.lanl.gov/abs/1010.0255}{{\tt arXiv:1010.0255}}].

\bibitem{Anderson:2011ty}
L.~B. Anderson, J.~Gray, A.~Lukas, and B.~Ovrut, {\it {The Atiyah Class and
  Complex Structure Stabilization in Heterotic Calabi-Yau Compactifications}},
  {\em JHEP} {\bf 10} (2011) 032,
  [\href{http://xxx.lanl.gov/abs/1107.5076}{{\tt arXiv:1107.5076}}].

\bibitem{Anderson:2011cza}
L.~B. Anderson, J.~Gray, A.~Lukas, and B.~Ovrut, {\it {Stabilizing All
  Geometric Moduli in Heterotic Calabi-Yau Vacua}},  {\em Phys.Rev.} {\bf D83}
  (2011) 106011, [\href{http://xxx.lanl.gov/abs/1102.0011}{{\tt
  arXiv:1102.0011}}].

\bibitem{Klaput:2012vv}
M.~Klaput, A.~Lukas, C.~Matti, and E.~E. Svanes, {\it {Moduli Stabilising in
  Heterotic Nearly K\"ahler Compactifications}},
  \href{http://xxx.lanl.gov/abs/1210.5933}{{\tt arXiv:1210.5933}}.

\bibitem{Klaput:2013nla}
M.~Klaput, A.~Lukas, and E.~E. Svanes, {\it {Heterotic Calabi-Yau
  Compactifications with Flux}},  {\em JHEP} {\bf 1309} (2013) 034,
  [\href{http://xxx.lanl.gov/abs/1305.0594}{{\tt arXiv:1305.0594}}].

\bibitem{Anderson:2014xha}
L.~B. Anderson, J.~Gray, and E.~Sharpe, {\it {Algebroids, Heterotic Moduli
  Spaces and the Strominger System}},  {\em JHEP} {\bf 1407} (2014) 037,
  [\href{http://xxx.lanl.gov/abs/1402.1532}{{\tt arXiv:1402.1532}}].

\bibitem{delaOssa:2014cia}
X.~de~la Ossa and E.~E. Svanes, {\it {Holomorphic Bundles and the Moduli Space
  of N=1 Heterotic Compactifications}},
  \href{http://xxx.lanl.gov/abs/1402.1725}{{\tt arXiv:1402.1725}}.

\bibitem{delaOssa:2014msa}
X.~de~la Ossa and E.~E. Svanes, {\it {Connections, Field Redefinitions and
  Heterotic Supergravity}},  \href{http://xxx.lanl.gov/abs/1409.3347}{{\tt
  arXiv:1409.3347}}.

\bibitem{Gurrieri:2002wz}
S.~Gurrieri, J.~Louis, A.~Micu, and D.~Waldram, {\it {Mirror Symmetry in
  Generalized Calabi-Yau Compactifications}},  {\em Nucl.Phys.} {\bf B654}
  (2003) 61--113, [\href{http://xxx.lanl.gov/abs/hep-th/0211102}{{\tt
  hep-th/0211102}}].

\bibitem{Gurrieri:2004dt}
S.~Gurrieri, A.~Lukas, and A.~Micu, {\it {Heterotic on Half-Flat}},  {\em
  Phys.Rev.} {\bf D70} (2004) 126009,
  [\href{http://xxx.lanl.gov/abs/hep-th/0408121}{{\tt hep-th/0408121}}].

\bibitem{Gurrieri:2007jg}
S.~Gurrieri, A.~Lukas, and A.~Micu, {\it {Heterotic String Compactifications on
  Half-Flat Manifolds. II.}},  {\em JHEP} {\bf 0712} (2007) 081,
  [\href{http://xxx.lanl.gov/abs/0709.1932}{{\tt arXiv:0709.1932}}].

\bibitem{Lust:1985be}
D.~Lust and G.~Zoupanos, {\it {Dimensional reduction of ten-dimensional E8
  gauge theory over a compact coset space S/R}},  {\em Phys.Lett.} {\bf B165}
  (1985) 309.

\bibitem{Lust:1986ix}
D.~L\"ust, {\it {Compactification of Ten-dimensional Superstring Theories Over
  Ricci Flat Coset Spaces}},  {\em Nucl.Phys.} {\bf B276} (1986) 220.

\bibitem{butruille}
J.-B. Butruille, {\it {Homogeneous nearly K\"{a}hler manifolds}},
  \href{http://xxx.lanl.gov/abs/0612655}{{\tt 0612655}}.

\bibitem{chatzistavrakidis2009reducing}
A.~Chatzistavrakidis, P.~Manousselis, and G.~Zoupanos, {\it Reducing the
  heterotic supergravity on nearly-k{\"a}hler coset spaces},  {\em Fortschritte
  der Physik} {\bf 57} (2009), no.~5-7 527--534.

\bibitem{Klaput:2011mz}
M.~Klaput, A.~Lukas, and C.~Matti, {\it {Bundles over Nearly-K\"ahler
  Homogeneous Spaces in Heterotic String Theory}},  {\em JHEP} {\bf 1109}
  (2011) 100, [\href{http://xxx.lanl.gov/abs/1107.3573}{{\tt
  arXiv:1107.3573}}].

\bibitem{Chatzistavrakidis:2012qb}
A.~Chatzistavrakidis, O.~Lechtenfeld, and A.~D. Popov, {\it {Nearly K\'ahler
  heterotic compactifications with fermion condensates}},  {\em JHEP} {\bf
  1204} (2012) 114, [\href{http://xxx.lanl.gov/abs/1202.1278}{{\tt
  arXiv:1202.1278}}].

\bibitem{Haupt:2014ufa}
A.~S. Haupt, O.~Lechtenfeld, and E.~T. Musaev, {\it {Order $\alpha'$ heterotic
  domain walls with warped nearly K\"ahler geometry}},
  \href{http://xxx.lanl.gov/abs/1409.0548}{{\tt arXiv:1409.0548}}.

\bibitem{Gallego:2008sv}
D.~Gallego and M.~Serone, {\it {Moduli Stabilization in non-Supersymmetric
  Minkowski Vacua with Anomalous U(1) Symmetry}},  {\em JHEP} {\bf 0808} (2008)
  025, [\href{http://xxx.lanl.gov/abs/0807.0190}{{\tt arXiv:0807.0190}}].

\bibitem{Blaszczyk:2014qoa}
M.~Blaszczyk, S.~Groot~Nibbelink, O.~Loukas, and S.~Ramos-Sanchez, {\it
  {Non-supersymmetric heterotic model building}},  {\em JHEP} {\bf 1410} (2014)
  119, [\href{http://xxx.lanl.gov/abs/1407.6362}{{\tt arXiv:1407.6362}}].

\bibitem{Kallosh:2014oja}
R.~Kallosh, A.~Linde, B.~Vercnocke, and T.~Wrase, {\it {Analytic Classes of
  Metastable de Sitter Vacua}},  {\em JHEP} {\bf 1410} (2014) 11,
  [\href{http://xxx.lanl.gov/abs/1406.4866}{{\tt arXiv:1406.4866}}].

\bibitem{Angelantonj:2014dia}
C.~Angelantonj, I.~Florakis, and M.~Tsulaia, {\it {Universality of Gauge
  Thresholds in Non-Supersymmetric Heterotic Vacua}},  {\em Phys.Lett.} {\bf
  B736} (2014) 365--370, [\href{http://xxx.lanl.gov/abs/1407.8023}{{\tt
  arXiv:1407.8023}}].

\bibitem{Breitenlohner1982197}
P.~Breitenlohner and D.~Freedman, {\it {Positive Energy in anti-De Sitter
  Backgrounds and Gauged Extended Supergravity}},  {\em Phys.Lett.} {\bf B115}
  (1982) 197.

\bibitem{micu2004heterotic}
A.~Micu, {\it Heterotic compactifications and nearly k{\"a}hler manifolds},
  {\em Physical Review D} {\bf 70} (2004), no.~12 126002.

\bibitem{KashaniPoor:2007tr}
A.-K. Kashani-Poor, {\it {Nearly K\"{a}hler Reduction}},  {\em JHEP} {\bf 0711}
  (2007) 026, [\href{http://xxx.lanl.gov/abs/0709.4482}{{\tt
  arXiv:0709.4482}}].

\bibitem{zoupanos2}
A.~Chatzistavrakidis and G.~Zoupanos, {\it Dimensional reduction of the
  heterotic string over nearly-kähler manifolds},  {\em Journal of High Energy
  Physics} {\bf 2009} (2009), no.~09 077.

\bibitem{Chatzistavrakidis:2009mh}
A.~Chatzistavrakidis and G.~Zoupanos, {\it {Dimensional Reduction of the
  Heterotic String over nearly-Kaehler manifolds}},  {\em JHEP} {\bf 0909}
  (2009) 077, [\href{http://xxx.lanl.gov/abs/0905.2398}{{\tt
  arXiv:0905.2398}}].

\bibitem{Lukas:2010mf}
A.~Lukas and C.~Matti, {\it {G-structures and Domain Walls in Heterotic
  Theories}},  {\em JHEP} {\bf 1101} (2011) 151,
  [\href{http://xxx.lanl.gov/abs/1005.5302}{{\tt arXiv:1005.5302}}].

\bibitem{CyrilThesis}
C.~Matti, {\it {Generalized Compactification in Heterotic String Theory}},
  \href{http://xxx.lanl.gov/abs/1204.3247}{{\tt arXiv:1204.3247}}.

\bibitem{Gray:2008zs}
J.~Gray, Y.-H. He, A.~Ilderton, and A.~Lukas, {\it {STRINGVACUA: A Mathematica
  Package for Studying Vacuum Configurations in String Phenomenology}},  {\em
  Comput.Phys.Commun.} {\bf 180} (2009) 107--119,
  [\href{http://xxx.lanl.gov/abs/0801.1508}{{\tt arXiv:0801.1508}}].

\bibitem{srednicki1985}
M.~Srednicki and S.~Theisen, {\it Supergravitational radiative corrections to
  the gauge hierarchy},  {\em Physical Review Letters} {\bf 54} (1985), no.~4
  278--280.

\bibitem{Gaillard:1993es}
M.~K. Gaillard and V.~Jain, {\it {Supergravity coupled to chiral matter at one
  loop}},  {\em Phys.Rev.} {\bf D49} (1994) 1951--1965,
  [\href{http://xxx.lanl.gov/abs/hep-th/9308090}{{\tt hep-th/9308090}}].

\bibitem{Gaillard:1996ms}
M.~K. Gaillard, V.~Jain, and K.~Saririan, {\it {Supergravity at one loop. 2:
  Chiral and Yang-Mills matter}},  {\em Phys.Rev.} {\bf D55} (1997) 883--924,
  [\href{http://xxx.lanl.gov/abs/hep-th/9606052}{{\tt hep-th/9606052}}].

\bibitem{Gaillard:1996hs}
M.~K. Gaillard, V.~Jain, and K.~Saririan, {\it {Supergravity coupled to chiral
  and Yang-Mills matter at one loop}},  {\em Phys.Lett.} {\bf B387} (1996)
  520--528, [\href{http://xxx.lanl.gov/abs/hep-th/9606135}{{\tt
  hep-th/9606135}}].

\bibitem{Witten:1996mz}
  E.~Witten,
  {\it Strong coupling expansion of Calabi-Yau compactification,}
  Nucl.\ Phys.\ B {\bf 471} (1996) 135,
  [\href{http://xxx.lanl.gov/abs/hep-th/9602070}{{\tt
  hep-th/9602070}}].
  
\bibitem{Buchbinder:2002ic}
  E.~I.~Buchbinder, R.~Donagi and B.~A.~Ovrut,
  {\it Superpotentials for vector bundle moduli,}
  Nucl.\ Phys.\ B {\bf 653} (2003) 400,
   [\href{http://xxx.lanl.gov/abs/hep-th/0205190}{{\tt
  hep-th/0205190}}].

\end{thebibliography}

\providecommand{\href}[2]{#2}\begingroup\raggedright\endgroup

\end{document}